\documentclass[reprint,amsmath,amssymb,aps,]{revtex4-2}

\usepackage{graphicx}
\usepackage{dcolumn}
\usepackage{bm}
\usepackage[]{hyperref}
\usepackage{color}
\usepackage{amsmath}

\begin{document}

\title{Analytical prediction of superradiant burst dynamics in extended atomic ensembles}
	
\author{Wei-Lin Chen,$^{1}$ Zi-Yu Liu,$^{1,2}$ Jiun-Shiuan Shiu,$^{1,2}$ and Yong-Fan Chen$^{1,2}$}
	
\email{yfchen@mail.ncku.edu.tw}
	
\affiliation{
	$^1$Department of Physics, National Cheng Kung University, Tainan 70101, Taiwan\\ 
	$^2$Center for Quantum Frontiers of Research $\&$ Technology, Tainan 70101, Taiwan
}
	
\date{November 17, 2026}
	
	
\begin{abstract}

We present an open-system quantum theory of forward superradiant bursts in extended two-level atomic ensembles using coupled Heisenberg--Langevin and Maxwell propagation equations. Starting from a fully inverted state with zero mean polarization, local dipole correlations initiate emission without a prescribed tipping angle or stochastic seed. Assuming a spatially uniform population inversion while retaining longitudinal field propagation and spatially dependent atomic polarization, we derive an analytical expression for the absolute photon emission rate into the selected forward optical channel. Coupled to self-consistent inversion dynamics, the expression quantitatively reproduces the pulse waveform and absolute peak rate from spatially resolved stochastic simulations for the sample parameters considered, up to an optical depth (OD) of approximately 200. At higher OD, longitudinal inversion gradients limit the approximation, while the peak rate trends toward Dicke-like quadratic scaling and the pulse width decreases approximately inversely with OD. The framework thus connects quantum initiation, propagation, and inversion depletion while delineating the limits of a spatially reduced description.

\end{abstract}
	
\maketitle
	
	
\section{Introduction}

Cooperative spontaneous emission is a fundamental manifestation of collective light--matter interaction. In Dicke's small-sample limit, the emitters behave as a collective dipole, producing a radiation burst whose peak intensity can scale quadratically with the atom number~\cite{Dicke1954}. Delayed superradiant bursts and subsequent ringing were observed experimentally in optically pumped HF gas~\cite{Skribanowitz1973}. In extended ensembles, however, spatial phase variations and field propagation become essential. Early theories established the roles of cooperative response and sample geometry~\cite{Arecchi1970,Rehler1971,Bonifacio1975a}, coupled field--matter propagation~\cite{MacGillivray1976,Polder1979}, and transverse diffraction~\cite{Gross1982}.

Burst initiation is particularly important for a fully inverted two-level ensemble. Because the initial mean polarization vanishes, the semiclassical Maxwell--Bloch equations remain at the zero-polarization solution unless a seed is introduced. Early quantum-statistical theories established the importance of atomic and field correlations~\cite{Bonifacio1971,Bonifacio1971a}, while quantum-based stochastic formulations represented the initiation of superfluorescence through fluctuating initial atomic polarizations~\cite{Haake1979}. Analytical quantum propagation theories subsequently described superfluorescence initiation within linearized treatments~\cite{Mostowski1984,Mostowski1986}. Recent approaches have developed quantum two-point correlation formulations~\cite{Benediktovitch2019} and stochastic Maxwell--Bloch equations with quantum-derived noise~\cite{Chuchurka2024}. Quantum many-body studies and experiments have explored fluctuation-driven collective emission~\cite{CardenasLopez2023,Liedl2024}.

Collective radiative dynamics have received renewed attention in superradiant lasers and active optical clocks, where collective atomic coherence provides a route toward stable optical sources with reduced sensitivity to cavity-frequency fluctuations~\cite{Meiser2009,Bohnet2012,Norcia2016}. Experiments have characterized spectra and phase synchronization~\cite{Norcia2018,Weiner2017}, while studies of atomic ensembles and arrays have examined directional emission, coherence buildup, and peak scaling~\cite{Ferioli2021,Robicheaux2021,RubiesBigorda2023}, including studies of second-order coherence~\cite{Ferioli2025,Bach2026}. Related developments include directional emission from single-excitation states, waveguide-mediated interactions, multimode photon generation, and many-body correlations~\cite{Scully2006,Goban2015,Paulisch2019,Zhang2025}, alongside disorder-robust waveguide superradiance and finite-size scaling~\cite{Zhang2026}. A key remaining question is whether a reduced analytical model can accurately predict the absolute forward photon emission rate throughout a superradiant burst while retaining longitudinal propagation and time-dependent inversion, and under what conditions longitudinal inversion gradients invalidate the approximation.

In this work, we develop an open-system quantum theory for the analytical prediction of forward superradiant burst dynamics in an extended two-level ensemble. Starting from a fully inverted state with zero mean polarization, local dipole correlations initiate emission in the operator-based formulation without a prescribed tipping angle or stochastic seed, while field-mediated interactions drive the cooperative buildup. Our central advancement is an analytical expression for the absolute forward emission rate derived by assuming a uniform population inversion while retaining longitudinal field propagation and spatially dependent atomic polarization. Coupled to self-consistent mean-inversion dynamics, this expression predicts the temporal waveform and absolute peak photon rate. Building on earlier analytical treatments of quantum initiation and propagation~\cite{Haake1979,Mostowski1984,Mostowski1986}, our formulation incorporates the evolving inversion into the propagation kernel, extending the analytical description to the depletion stage of the cooperative burst. For optical depths (OD) up to $\sim 200$, these predictions agree quantitatively with spatially resolved stochastic simulations initialized with matching local dipole correlations.

The analytical and spatially resolved treatments further reveal distinct OD-dependent regimes. Within the independent-decay approximation, the peak time formally approaches $\ln 2$ times the single-atom radiative lifetime from below, with a leading $1/\sqrt{\mathrm{OD}}$ correction. At higher OD, longitudinal inversion gradients reverse this peak-time trend, exposing the limits of the uniform-inversion approximation. The spatially resolved simulations also reveal a nonmonotonic local peak-rate exponent exceeding two at intermediate OD, followed by trends toward Dicke-like quadratic peak-rate growth and inverse-OD pulse narrowing at the highest OD studied. In the low-excitation limit, the same framework yields an OD-enhanced short-time forward-emission decay.


\section{Open-system formulation and quantum-fluctuation initiation}
\label{sec:open_system}

\subsection{One-dimensional open-system model}
\label{subsec:1D_model}

We consider an extended one-dimensional (1D) ensemble of two-level atoms occupying $0\leq z\leq L$ and resonantly coupled to a selected forward-propagating optical channel~\cite{MacGillivray1976,Polder1979,Gross1982}. The transverse profile of the selected channel is represented by an effective mode area $\mathcal{A}$, while the remaining radiative modes are traced out and represented by radiative damping and Langevin forces. In the rotating frame and under the electric-dipole and rotating-wave approximations, the interaction Hamiltonian of the explicitly retained forward channel is
\begin{equation}
	\label{eq:hamiltonian}
	\begin{split}
		\hat{\mathcal{H}}_{\mathrm{int}}
		={}&-\frac{\hbar N}{L}\int_{0}^{L}\mathrm{d}z\,
		\Bigl[
		g\hat{\sigma}_{21}(z,t)\hat{a}(z,t)\\
		&\qquad\qquad
		+g^*\hat{a}^{\dagger}(z,t)
		\hat{\sigma}_{12}(z,t)
		\Bigr].
	\end{split}
\end{equation}
Here, $N$ is the number of atoms within the effective mode volume, $g$ is the single-atom coupling coefficient to the selected forward channel, and $\hat{a}(z,t)$ is the slowly varying annihilation operator of the forward field, normalized such that $[\hat{a}(z,t), \hat{a}^{\dagger}(z',t)] = L\delta(z-z')$.

The dimensionless operator $\hat{\sigma}_{\mu\nu}(z,t)$ denotes the coarse-grained local collective atomic transition, normalized so that $N\hat{\sigma}_{\mu\nu}(z,t)/L$ is the corresponding transition density. Defining $\hat{\sigma}_{3}=\hat{\sigma}_{22}-\hat{\sigma}_{11}$, the relevant local commutation relation is
\begin{equation}
	\label{eq:commutation}
	\bigl[
	\hat{\sigma}_{12}(z,t),
	\hat{\sigma}_{21}(z',t)
	\bigr]
	=
	-\frac{L}{N}\delta(z-z')
	\hat{\sigma}_{3}(z,t).
\end{equation}
Within the slowly varying envelope approximation, we introduce the retarded time $\tau=t-z/c$. The coupled Maxwell and Heisenberg--Langevin equations then read
\begin{equation}
	\label{eq:wave_eq}
	\frac{\partial}{\partial z}
	\hat{a}(z,\tau)
	=
	i\frac{g^*N}{c}
	\hat{\sigma}_{12}(z,\tau),
\end{equation}
\begin{equation}
	\label{eq:hle_pol}
	\begin{split}
		\frac{\partial}{\partial\tau}
		\hat{\sigma}_{12}(z,\tau)
		={}&
		-\frac{\Gamma_{21}}{2}
		\hat{\sigma}_{12}(z,\tau)\\
		&-ig\hat{\sigma}_{3}(z,\tau)
		\hat{a}(z,\tau)
		+\hat{F}_{12}(z,\tau),
	\end{split}
\end{equation}
and
\begin{equation}
	\label{eq:hle_inv}
	\begin{split}
		\frac{\partial}{\partial\tau}
		\hat{\sigma}_{3}(z,\tau)
		={}&
		-\Gamma_{21}
		\bigl[
		\hat{\sigma}_{3}(z,\tau)+\hat{I}
		\bigr]\\
		&+2i
		\Bigl[
		g\hat{\sigma}_{21}(z,\tau)
		\hat{a}(z,\tau)\\
		&\qquad\qquad
		-g^*\hat{a}^{\dagger}(z,\tau)
		\hat{\sigma}_{12}(z,\tau)
		\Bigr]
		+\hat{F}_{3}(z,\tau).
	\end{split}
\end{equation}
The equation for $\hat{\sigma}_{21}$ is the Hermitian conjugate of Eq.~(\ref{eq:hle_pol}). Here, $\Gamma_{21}$ is the single-atom radiative population decay rate associated with the eliminated reservoir modes, while the zero-mean Langevin operators $\hat{F}_{12}$ and $\hat{F}_{3}$ describe reservoir fluctuations and preserve the atomic operator commutation relations under dissipative evolution~\cite{Gardiner2004}. For the paraxial geometry considered here, the fraction of single-atom spontaneous emission coupled into the explicitly retained forward channel is negligible ($\sim\lambda^2/\mathcal{A}\ll 1$), and $\Gamma_{21}$ can therefore be identified with the free-space radiative population decay rate.

The primary observable considered below is the forward photon emission rate at the sample exit $z=L$, defined from the normally ordered field correlation as
\begin{equation}
	\label{eq:emission_rate}
	R(\tau)
	=
	\frac{c}{L}
	\bigl\langle
	\hat{a}^{\dagger}(L,\tau)\hat{a}(L,\tau)
	\bigr\rangle.
\end{equation}
This rate refers to the selected forward optical channel rather than the total emission over all transverse modes. Direct comparison with experimental photon counts therefore requires accounting for the paraxial mode overlap and collection efficiency. At the ensemble entrance, $\hat{a}(0,\tau)=\hat{a}_{\mathrm{in}}(\tau)$ denotes the incident vacuum field, whose normally ordered photon flux vanishes.

In the numerical calculations, we model the atomic medium as an effective two-level system based on the $D_2$ line of laser-cooled $^{87}\mathrm{Rb}$. A possible realization is the closed $\sigma^+$ cycling transition between the stretched states $|1\rangle=|5S_{1/2},F=2,m_F=+2\rangle$ and $|2\rangle=|5P_{3/2},F'=3,m_{F'}=+3\rangle$. We assume ideal preparation of the fully inverted ensemble and neglect coupling to other hyperfine and Zeeman sublevels. We take the resonant wavelength and radiative population decay rate to be $\lambda=780~\mathrm{nm}$ and $\Gamma_{21}\simeq2\pi\times6.067~\mathrm{MHz}$, respectively~\cite{SteckRb87}. The sample length is taken to be $L=4~\mathrm{mm}$, and a representative effective transverse mode area is chosen as $\mathcal{A}=\pi(d/2)^2$ with an effective mode diameter $d=1~\mathrm{mm}$. This transverse scale fixes the corresponding effective atom number used in the numerical benchmark but does not constitute an independently determined parameter of the 1D model. For the chosen parameters, $\mathcal{A}/(\lambda L)\gg1$, so transverse diffraction of the selected mode over the sample length is negligible within the present 1D propagation treatment.

The retarded-time coordinate removes the explicit free-space transit term while retaining the spatial evolution in Eq.~(\ref{eq:wave_eq}). For $L=4~\mathrm{mm}$, the free-space transit time is $L/c\simeq13.3~\mathrm{ps}$. We then relabel the retarded-time variable $\tau$ as $t$ when discussing the emission dynamics. This relabeling is purely notational and does not remove longitudinal field propagation or the spatial dependence of the atomic polarization~\cite{Gross1982,Polder1979}.

The uniform ensemble has an atomic number density $n=N/(\mathcal{A}L)$, giving $\mathrm{OD}=n\sigma_{0}L=N\sigma_{0}/\mathcal{A}$, where $\sigma_{0}=3\lambda^{2}/(2\pi)$ is the resonant scattering cross section of the effective closed two-level transition. OD is defined for the corresponding ground-state medium and serves as a coupling parameter for the initially inverted ensemble. At fixed $L$ and $\mathcal{A}$, varying OD is therefore equivalent to varying the atom number, $N=\mathrm{OD}\,\mathcal{A}/\sigma_{0}$. Conversely, at fixed OD the 1D model determines the column density $N/\mathcal{A}$ rather than the absolute atom number and transverse area separately. With the field normalization adopted above, the forward collective coupling is related to OD by
\begin{equation}
	\label{eq:g_od_relation}
	\frac{|g|^{2}N}{c}
	=
	\frac{\mathrm{OD}\,\Gamma_{21}}{4L}.
\end{equation}
Thus, for fixed OD and $L$, changes in $\mathcal{A}$ are accompanied by corresponding changes in $N$ and $g$, while the collective coupling $|g|^2N$ remains unchanged within the 1D continuum description. We use this relation below to parameterize the collective interaction strength by OD. Although the forward collective coupling is determined by OD, the total atom number can still affect the emission dynamics through the depletion of the atomic population inversion, as discussed in Sec.~\ref{subsec:cooperative_peak_time}.


\subsection{Quantum-fluctuation initiation}
\label{subsec:quantum_initiation}

The ensemble is initially prepared in the fully inverted product state $|2\rangle^{\otimes N}$, for which the local mean polarization vanishes, $\langle\hat{\sigma}_{12}(z,0)\rangle=0$. In the absence of a coherent incident field or stochastic seed, the semiclassical Maxwell--Bloch equations remain at the zero-polarization and zero-field solution. By contrast, the operator description permits a vanishing mean polarization to coexist with a nonzero equal-time two-point dipole correlation. For two longitudinal positions $z_1$ and $z_2$ within the ensemble, the product state and the local operator algebra give
\begin{equation}
	\label{eq:quantum_seed}
	\begin{split}
		&\left\langle
		\hat{\sigma}_{21}(z_1,0)
		\hat{\sigma}_{12}(z_2,0)
		\right\rangle\\
		&\qquad=
		\frac{L}{N}\delta(z_1-z_2)
		\left\langle
		\hat{\sigma}_{22}(z_1,0)
		\right\rangle .
	\end{split}
\end{equation}
For a fully inverted ensemble, $\langle\hat{\sigma}_{22}(z_1,0)\rangle=1$.
This $\delta$-correlated form shows that the initial dipole correlations are local, with no pre-existing macroscopic phase coherence imposed across the ensemble.
Through the subsequent collective light--matter interaction, this microscopic atomic correlation drives the buildup of nonzero normally ordered forward-field moments.

Assuming independent local atomic decay into a zero-temperature Markov reservoir, the corresponding Langevin noise correlations under the coarse-grained atomic normalization evaluate to
\begin{equation}
	\label{eq:langevin_correlations}
	\begin{aligned}
		\left\langle
		\hat{F}_{21}(z,t)\hat{F}_{12}(z',t')
		\right\rangle
		&=0,\\
		\left\langle
		\hat{F}_{12}(z,t)\hat{F}_{21}(z',t')
		\right\rangle
		&=
		\frac{L}{N}\Gamma_{21}
		\delta(z-z')\delta(t-t').
	\end{aligned}
\end{equation}
Importantly, the vanishing of the normally ordered correlation, $\langle\hat{F}_{21}\hat{F}_{12}\rangle=0$, implies that the reservoir noise does not provide a direct normally ordered source term for the forward photon flux $R(t)$, while the reverse-ordered correlation remains nonzero and is required to preserve operator commutation relations under dissipation. Accordingly, the nonzero initial contribution to the forward emission originates from the atomic dipole correlation in Eq.~(\ref{eq:quantum_seed}). No prescribed tipping angle or stochastic seed is therefore introduced in the operator-based analytical treatment. We refer to the resulting dynamics as quantum-fluctuation-initiated emission.


\section{Propagation dynamics and limits of the spatially uniform inversion approximation}
\label{sec:spatial_dynamics}

\subsection{Longitudinal inversion profiles}
\label{subsec:inversion_profiles}

To assess the longitudinal homogeneity of the population inversion, we use a spatially resolved numerical benchmark based on the 1D formulation introduced in Sec.~\ref{sec:open_system}, retaining both forward-field propagation and the local atomic dynamics~\cite{MacGillivray1976,Polder1979}. We define the local inversion expectation value as $\sigma_3(z,t) \equiv \langle \hat{\sigma}_3(z,t) \rangle$. In the numerical calculations, the medium is discretized with a spatial resolution $\Delta z = 0.004~\mathrm{mm}$ and an OD-dependent time step $\Delta t = \tau_R/250$, where $\tau_R = (\mathrm{OD}\,\Gamma_{21})^{-1}$ sets the cooperative timescale used to define the temporal resolution. Numerical convergence of the calculated observables was verified across the studied OD range with respect to spatial and temporal discretization and the number of stochastic realizations.

In the spatially resolved benchmark calculation, quantum initiation is modeled through a stochastic initial polarization seed following established theoretical treatments of cooperative emission~\cite{MacGillivray1976,Polder1979,Haake1979}, with its amplitude chosen to reproduce the local second-order dipole correlation in Eq.~(\ref{eq:quantum_seed}). Specifically, for each realization of the ensemble, the local atomic coherence is initialized as $\sigma_{12}(z,0) = \frac{1}{2} \sin\theta_0 \, e^{i\phi(z)} \approx \frac{1}{2}\theta_0 e^{i\phi(z)}$, where $\theta_0 \simeq 2/\sqrt{N_{\mathrm{cell}}}$ defines the local tipping angle scale and $N_{\mathrm{cell}} = N \Delta z / L$ is the atom number per spatial cell. The phase $\phi(z)$ is drawn independently and uniformly from $[0,2\pi)$ in each cell of size $\Delta z$. The initial inversion is accordingly set to $\sigma_3(z,0) = \sqrt{1 - 4|\sigma_{12}(z,0)|^2} = \cos\theta_0 \simeq 1$. For spatial cells indexed by $j$ and $k$, the stochastic ensemble averages satisfy $\langle\sigma_{12,j}(0)\rangle_{\mathrm{st}}=0$ and $\langle\sigma_{21,j}(0)\sigma_{12,k}(0)\rangle_{\mathrm{st}}=\frac{1}{4}\sin^2\theta_0\,\delta_{jk}\simeq\delta_{jk}/N_{\mathrm{cell}}$, reproducing the discrete form of Eq.~(\ref{eq:quantum_seed}). Macroscopic observables, including the forward photon emission rate at the sample exit $z=L$, are obtained by averaging over $M = 5000$ independent stochastic trajectories. This effective seed is used only to initialize the spatially resolved numerical benchmark and is not introduced in the operator-based analytical treatment developed below.

\begin{figure}[t]
	\centering
	\includegraphics[width=\columnwidth]{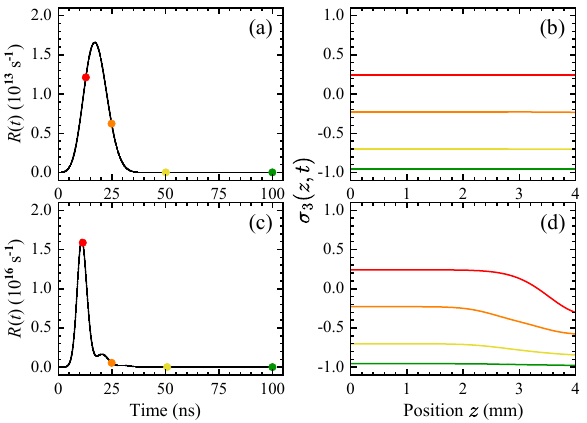}
	\caption{Spatially resolved stochastic Maxwell--Bloch simulations of the forward emission and local inversion dynamics. (a) Forward photon emission rate $R(t)$ at the sample exit $z=L$ for $\mathrm{OD}=200$. (b) Corresponding longitudinal profiles of the local inversion $\sigma_3(z,t)$. (c) Forward photon emission rate for $\mathrm{OD}=500$. (d) Corresponding longitudinal profiles, showing a pronounced inversion gradient with faster depletion near the output end $z=L$. The red, orange, yellow, and green circles in (a) and (c) mark $t=12.5$, $25$, $50$, and $100~\mathrm{ns}$, respectively, with the inversion profiles in (b) and (d) shown in the corresponding colors.}
	\label{fig:fig1}
\end{figure}

Figure~\ref{fig:fig1} compares the forward emission dynamics and longitudinal inversion profiles for $\mathrm{OD}=200$ and $500$. At $\mathrm{OD}=200$, the ensemble produces a delayed emission pulse, while $\sigma_3(z,t)$ remains nearly independent of position during most of the collective decay [Figs.~\ref{fig:fig1}(a) and \ref{fig:fig1}(b)]. This behavior motivates a spatially uniform inversion approximation in this regime. Importantly, this approximation is applied only to the inversion and does not require the atomic polarization or propagating field to be spatially uniform.

At $\mathrm{OD}=500$, the emission occurs earlier than for $\mathrm{OD}=200$ and reaches a substantially larger peak. The local inversion simultaneously develops a pronounced longitudinal gradient and is depleted more rapidly near the output end of the sample, $z=L$ [Figs.~\ref{fig:fig1}(c) and \ref{fig:fig1}(d)]. This spatial dependence is consistent with the buildup of the forward field during propagation, which enhances collective depletion toward the output end. The smaller secondary peaks following the main burst in Fig.~\ref{fig:fig1}(c) are consistent with coherent ringing associated with reabsorption and reemission during propagation in an optically thick medium~\cite{Gross1982}. The contrasting behaviors at $\mathrm{OD}=200$ and $500$ motivate the analytical treatment developed below and illustrate the growing limitations of a spatially uniform inversion description at high OD.

	
\subsection{Analytical propagation solution for spatially uniform inversion}
\label{subsec:uniform_inversion_solution}

Motivated by the nearly uniform inversion profile at $\mathrm{OD}=200$ in Fig.~\ref{fig:fig1}(b), we introduce the spatially uniform inversion approximation. We replace the local inversion operator by the spatially averaged inversion expectation value,
\begin{equation}
	\label{eq:uniform_inversion}
	\hat{\sigma}_3(z,t)
	\rightarrow
	\bar{\sigma}_3(t)\hat{I},
	\qquad
	\bar{\sigma}_3(t)
	\equiv
	\left\langle
	\frac{1}{L}
	\int_{0}^{L}
	\hat{\sigma}_3(z,t)\,\mathrm{d}z
	\right\rangle .
\end{equation}
This approximation represents a spatial homogenization of the population inversion through its macroscopic mean trajectory $\bar{\sigma}_3(t)$, while retaining the time-dependent depletion of the inversion and capturing longitudinal field propagation and spatial variations in atomic polarization. For any prescribed $\bar{\sigma}_3(t)$, the coupled field--polarization sector reduces to a linear operator propagation problem.

To obtain the propagation solution, we introduce the unilateral spatial Laplace transforms
\begin{equation}
	\label{eq:spatial_Laplace_transforms}
	\begin{aligned}
		\hat{\Sigma}_{12}(s,t)
		&=
		\int_{0}^{\infty}
		\hat{\sigma}_{12}(z,t)e^{-sz}\,\mathrm{d}z,\\
		\hat{A}(s,t)
		&=
		\int_{0}^{\infty}
		\hat{a}(z,t)e^{-sz}\,\mathrm{d}z .
	\end{aligned}
\end{equation}
Here, $s$ is the complex Laplace variable. As an auxiliary construction, the propagation problem is formally continued to $z\geq0$, while the inverse transform is evaluated only in the physical region $0\leq z\leq L$. Because propagation is first order and unidirectional, the solution at a given position depends only on upstream sources and is therefore independent of the continuation beyond $L$. The Langevin operator is transformed in the same manner, and its spatial Laplace transform is denoted by $\hat{f}_{12}(s,t)$. Applying the transform to Eq.~(\ref{eq:wave_eq}) gives
\begin{equation}
	\label{eq:LT_wave}
	s\hat{A}(s,t)
	-
	\hat{a}_{\mathrm{in}}(t)
	=
	i\frac{g^*N}{c}
	\hat{\Sigma}_{12}(s,t).
\end{equation}
We separate the transformed field and polarization into vacuum-input and atom-generated components. Under the spatially uniform inversion approximation, the resulting linear operator dynamics allow the transformed polarization to be decomposed as $\hat{\Sigma}_{12}(s,t)=\hat{\Sigma}_{12}^{(\mathrm{at})}(s,t)+\hat{\Sigma}_{12}^{(\mathrm{in})}(s,t)$. The atomic-source component is driven by the initial atomic polarization alongside reservoir Langevin forces, whereas the input-driven component represents the linear response to the incident vacuum field. Since the latter is linear in the annihilation operators of the incident vacuum, it yields zero contribution to the normally ordered output photon flux. Cross correlations with the atomic-source component likewise vanish for initially uncorrelated field, atomic, and reservoir states. We therefore focus on the atom-generated component,
\begin{equation}
	\label{eq:LT_atom_generated_field}
	\hat{A}_{\mathrm{at}}(s,t)
	=
	i\frac{g^*N}{cs}
	\hat{\Sigma}_{12}^{(\mathrm{at})}(s,t).
\end{equation}

After applying the spatial Laplace transform to Eq.~(\ref{eq:hle_pol}), we substitute Eq.~(\ref{eq:LT_atom_generated_field}) into the transformed polarization equation and obtain the evolution equation for the atomic-source polarization,
\begin{equation}
	\label{eq:decoupled_ODE}
	\begin{split}
		\frac{\partial}{\partial t}
		\hat{\Sigma}_{12}^{(\mathrm{at})}(s,t)
		={}&
		\left[
		\frac{|g|^2N}{cs}\bar{\sigma}_3(t)
		-
		\frac{\Gamma_{21}}{2}
		\right]
		\hat{\Sigma}_{12}^{(\mathrm{at})}(s,t)\\
		&+
		\hat{f}_{12}(s,t).
	\end{split}
\end{equation}
The $\bar{\sigma}_3(t)$-dependent term describes the inversion-dependent collective coupling and provides collective gain for positive inversion, whereas $\hat{f}_{12}(s,t)$ accounts for fluctuations associated with the radiative reservoir~\cite{Gardiner2004}. For negative inversion, the same collective term changes sign and describes absorptive rather than amplifying propagation. The time-dependent inversion thus captures the crossover between collective gain and resonant absorption. Using Eq.~(\ref{eq:g_od_relation}), we define the integrated inversion parameter
\begin{equation}
	\label{eq:Phi_definition}
	\Phi(t)
	=
	\frac{\mathrm{OD}\Gamma_{21}}{4}
	\int_{0}^{t}
	\bar{\sigma}_3(t')\,\mathrm{d}t' .
\end{equation}
The evolution from time $t'$ to $t$ is described by the propagator
\begin{equation}
	\label{eq:polarization_propagator}
	\mathcal{G}(s;t,t')
	=
	\exp\left[
	-\frac{\Gamma_{21}}{2}(t-t')
	+
	\frac{\Phi(t)-\Phi(t')}{sL}
	\right].
\end{equation}
The formal atomic-source polarization solution can then be written compactly as
\begin{equation}
	\label{eq:formal_polarization_solution}
	\begin{split}
		\hat{\Sigma}_{12}^{(\mathrm{at})}(s,t)
		={}&
		\mathcal{G}(s;t,0)
		\hat{\Sigma}_{12}^{(\mathrm{at})}(s,0)\\
		&+
		\int_{0}^{t}\mathrm{d}t'\,
		\mathcal{G}(s;t,t')
		\hat{f}_{12}(s,t').
	\end{split}
\end{equation}
The first term evolves the initial atomic polarization fluctuations of the inverted ensemble, whereas the second contains the Langevin contribution accumulated during the subsequent open-system evolution~\cite{Gardiner2004}. In the propagator, $\Phi(t)-\Phi(t')$ is the integrated inversion-dependent collective contribution over the interval from $t'$ to $t$. Positive inversion amplifies the atomic polarization and forward field, while the accompanying collective emission depletes the inversion and progressively weakens the gain.

The atom-generated physical-space field is obtained by performing the inverse Laplace transform and evaluating the result at the sample output $z=L$. Using $\mathcal{L}^{-1}_{s\rightarrow z}\{s^{-1}\exp(b/s)\}=I_0(2\sqrt{bz})$, together with the initial dipole correlation in Eq.~(\ref{eq:quantum_seed}), we evaluate the forward photon flux $R(t)$ defined in Sec.~\ref{subsec:1D_model}. Because the normally ordered reservoir correlation vanishes [Eq.~(\ref{eq:langevin_correlations})], the accumulated Langevin term gives no direct contribution to $R(t)$. These Langevin operators nevertheless remain indispensable for preserving the atomic commutation relations under dissipative open-system dynamics~\cite{Gardiner2004}. Mapping the remaining spatial integral onto $0\leq\xi\leq1$ and using $\int_{0}^{1}\mathrm{d}\xi\,I_0^2(2\sqrt{\Phi \xi})=I_0^2(2\sqrt{\Phi})-I_1^2(2\sqrt{\Phi})$, we obtain the forward photon emission rate
\begin{equation}
	\label{eq:uniform_inversion_solution}
	\begin{split}
		R(t)
		={}&
		\frac{\mathrm{OD}\Gamma_{21}}{4}
		e^{-\Gamma_{21}t}\\
		&\times
		\left[
		I_0^2\left(2\sqrt{\Phi(t)}\right)
		-
		I_1^2\left(2\sqrt{\Phi(t)}\right)
		\right].
	\end{split}
\end{equation}
Here, $I_n$ denotes the modified Bessel function of the first kind. For $\Phi(t)<0$, the expression is evaluated by analytic continuation and remains real and nonnegative. Since $\Phi(0)=0$, $I_0(0)=1$ and $I_1(0)=0$, Eq.~(\ref{eq:uniform_inversion_solution}) gives the initial forward emission rate $R(0)=\mathrm{OD}\Gamma_{21}/4$. Within the operator formulation, this initial value is fixed by the local dipole correlation in Eq.~(\ref{eq:quantum_seed}) together with the OD parametrization of the forward collective coupling in Eq.~(\ref{eq:g_od_relation}), and is therefore independent of the spatially uniform inversion approximation. The forward emission is governed by the integrated inversion history $\Phi(t)$, which requires $\bar{\sigma}_3(t)$ to be determined self-consistently to close the dynamics.

	
\subsection{Cooperative dynamics and peak-time dependence on OD}
\label{subsec:cooperative_peak_time}

\begin{figure}[t]
	\centering
	\includegraphics[width=\columnwidth]{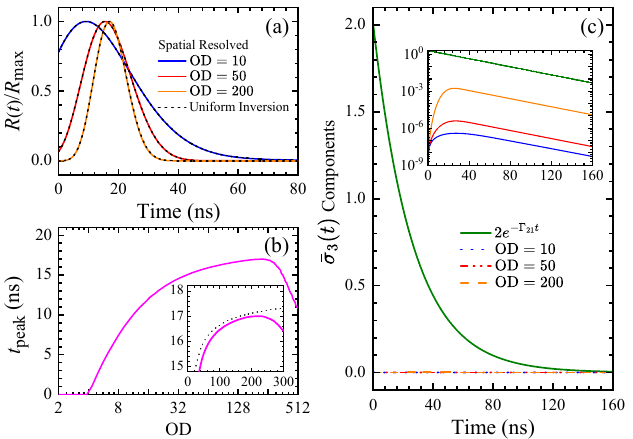}
	\caption{Normalized photon emission profiles, peak-time scaling, and inversion-depletion dynamics. (a) Normalized forward emission rate $R(t)/R_{\mathrm{max}}$ for $\mathrm{OD} = 10, 50,$ and $200$, with solid and dashed lines denoting the spatially resolved and uniform-inversion results, respectively. (b) Peak time $t_{\mathrm{peak}}$ versus OD on a $\log_2$ scale extending beyond $\mathrm{OD} = 500$. The inset compares the spatially resolved result (solid line) with the analytical estimate (dashed line). (c) Single-atom decay term $2e^{-\Gamma_{21}t}$ and collective-depletion term $(2/N)I_{\mathrm{coll}}(t)$ for the same OD values, with $I_{\mathrm{coll}}$ defined in Eq.~(\ref{eq:I_coll_definition}). The constant term $-1$ in Eq.~(\ref{eq:inversion_decomposition}) is omitted for clarity, and the inset uses a logarithmic vertical scale.}
	\label{fig:fig2}
\end{figure}

To assess the spatially uniform inversion approximation, we compare the normalized photon emission profiles $R(t)/R_{\mathrm{max}}$ for $\mathrm{OD}=10$, $50$, and $200$. As shown in Fig.~\ref{fig:fig2}(a), the uniform-inversion solution closely reproduces the spatially resolved normalized profiles over this OD range. As OD increases, the normalized emission pulse becomes narrower. As OD decreases, the collective buildup beyond the initial forward emission becomes progressively weaker, causing the emission maximum to shift toward $t=0$. In the low-OD limit, $R_{\mathrm{max}}$ therefore approaches the initial value $R(0)=\mathrm{OD}\Gamma_{21}/4$ given by Eq.~(\ref{eq:uniform_inversion_solution}). This comparison tests the temporal pulse shape rather than the absolute forward photon emission rate, which is examined below.

The peak emission time $t_{\mathrm{peak}}$ exhibits a nonmonotonic dependence on OD. The spatially resolved result in Fig.~\ref{fig:fig2}(b) shows that $t_{\mathrm{peak}}$ increases as OD is raised from $10$ to $200$, with a progressively weaker OD dependence. The inset compares this result with the analytical estimate derived in the following subsection, which additionally approximates the inversion by its independent-decay trajectory. Over the moderate-OD range, the analytical estimate reproduces the overall OD dependence of the spatially resolved result, although a visible quantitative deviation remains.

The discrepancy grows at higher OD, reflecting the limitations of the analytical estimate, which assumes both spatially uniform inversion and independent-decay dynamics. Figure~\ref{fig:fig1}(d) independently shows a pronounced longitudinal inversion gradient at $\mathrm{OD}=500$, indicating the breakdown of the spatially uniform inversion approximation. At still higher OD, the spatially resolved curve turns around and $t_{\mathrm{peak}}$ decreases [Fig.~\ref{fig:fig2}(b)]. This turnaround is consistent with increasingly rapid inversion depletion near the output end and the resulting shortening of the effective amplification period. To understand the moderate-OD peak-time behavior, we next examine the population dynamics governing the collective amplification.

Taking the expectation value and spatial average of Eq.~(\ref{eq:hle_inv}), using Eq.~(\ref{eq:wave_eq}) and the zero mean of the Langevin force, gives a population balance that separates single-atom radiative decay from depletion associated with forward-channel emission. Integrating this relation for the initially inverted ensemble yields
\begin{equation}
	\label{eq:inversion_decomposition}
	\bar{\sigma}_3(t) = \left( 2e^{-\Gamma_{21}t} - 1 \right) - \frac{2}{N} I_{\mathrm{coll}}(t),
\end{equation}
where $2e^{-\Gamma_{21}t}-1$ is the independent-decay inversion, while the second term describes the additional depletion due to forward collective emission, with
\begin{equation}
	\label{eq:I_coll_definition}
	I_{\mathrm{coll}}(t) = \int_{0}^{t} e^{-\Gamma_{21}(t-t')} R(t')\,\mathrm{d}t' .
\end{equation}
The factor $2/N$ implies that ensembles with the same OD but different effective mode areas need not exhibit identical inversion depletion or integrated forward photon numbers. Figure~\ref{fig:fig2}(c) and its inset compare the single-atom term $2e^{-\Gamma_{21}t}$ with the collective-depletion term $(2/N)I_{\mathrm{coll}}(t)$. For the moderate-OD cases shown, the collective contribution increases with OD but remains small relative to the single-atom term through most of the buildup. This factor converts the weighted forward photon number into the inversion change per atom, while the exponential kernel in Eq.~(\ref{eq:I_coll_definition}) accounts for the single-atom radiative decay. Consequently, the spatially averaged inversion remains close to the independent-decay trajectory $\bar{\sigma}_3^{(0)}(t)\equiv2e^{-\Gamma_{21}t}-1$ through most of the buildup in this regime.

This decomposition clarifies the moderate-OD behavior of $t_{\mathrm{peak}}$. Positive inversion provides gain for the atomic polarization and forward field, while the radiative reservoir continuously depletes the inversion on the single-atom timescale $\Gamma_{21}^{-1}$. Increasing OD strengthens the collective buildup and narrows the emission pulse, but the available gain window remains governed largely by the reservoir-driven population decay. This interplay between collective amplification and inversion depletion motivates the analytical peak-time treatment below.

	
\subsection{Analytical peak-time asymptotics}
\label{subsec:analytical_peak_evaluation}

To obtain an analytical estimate of the peak time, we further approximate the spatially uniform inversion by the independent-decay trajectory $\bar{\sigma}_3(t)\simeq\bar{\sigma}_3^{(0)}(t)=2e^{-\Gamma_{21}t}-1$. The integrated inversion parameter entering the collective propagator becomes $\Phi^{(0)}(t)=(\mathrm{OD}/4)[2(1-e^{-\Gamma_{21}t})-\Gamma_{21}t]$. This replacement neglects the feedback of forward collective emission on the spatially averaged inversion while retaining the inversion-dependent collective coupling in the polarization and field propagation through the propagator determined by $\Phi^{(0)}(t)$. Differentiating the emission rate in Eq.~(\ref{eq:uniform_inversion_solution}) and using $\mathrm{d}[I_0^2(2\sqrt{\Phi})-I_1^2(2\sqrt{\Phi})]/\mathrm{d}\Phi=I_1^2(2\sqrt{\Phi})/\Phi$, the condition $\mathrm{d}R(t)/\mathrm{d}t=0$ at $t=t_{\mathrm{peak}}$ is required for a delayed interior maximum with $t_{\mathrm{peak}}>0$. At sufficiently low OD where collective gain cannot overcome single-atom decay, the emission rate decays monotonically from $t=0$, so the global maximum resides at the boundary $t=0$. In the cooperative regime where a delayed burst emerges, this interior stationarity condition yields
\begin{equation}
	\label{eq:transcendental_peak}
	\frac{\mathrm{OD}}{4}
	\bar{\sigma}_3^{(0)}(t_{\mathrm{peak}})
	\frac{
		I_1^2\left(2\sqrt{\Phi_p^{(0)}}\right)
	}{
		I_0^2\left(2\sqrt{\Phi_p^{(0)}}\right)
		-
		I_1^2\left(2\sqrt{\Phi_p^{(0)}}\right)
	}
	=
	\Phi_p^{(0)} .
\end{equation}
Here, $u_p=\Gamma_{21}t_{\mathrm{peak}}$, $\bar{\sigma}_3^{(0)}(t_{\mathrm{peak}})=2e^{-u_p}-1$, and $\Phi_p^{(0)}=\Phi^{(0)}(t_{\mathrm{peak}})=(\mathrm{OD}/4)[2(1-e^{-u_p})-u_p]$. Equation~(\ref{eq:transcendental_peak}) is the exact stationarity condition for the analytical emission rate under the independent-decay approximation for a delayed peak with $\Phi_p^{(0)}>0$.

\begin{figure}[t]
	\centering
	\includegraphics[width=\columnwidth]{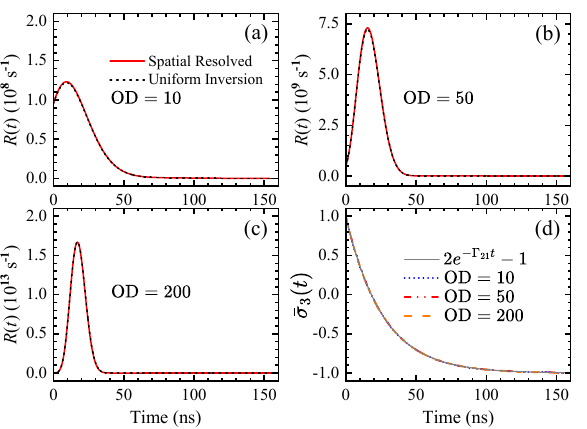}
	\caption{Comparison between spatially resolved calculations and the uniform-inversion approximation. In (a)--(c), solid curves show the spatially resolved calculation and dotted curves the spatially uniform inversion approximation. Forward photon emission rate $R(t)$ at (a) $\mathrm{OD}=10$, (b) $\mathrm{OD}=50$, and (c) $\mathrm{OD}=200$. (d) Inversion $\bar{\sigma}_3(t)$ predicted by the spatially uniform inversion approximation for $\mathrm{OD}=10$, $50$, and $200$ (colored dashed/dotted curves), together with the independent-decay reference $2e^{-\Gamma_{21}t}-1$ (solid gray line).}
	\label{fig:fig3}
\end{figure}

For sufficiently large $\Phi_p^{(0)}$, the modified Bessel functions satisfy
\begin{equation}
	\left[
	\frac{I_0(2\sqrt{\Phi_p^{(0)}})}
	{I_1(2\sqrt{\Phi_p^{(0)}})}
	\right]^2
	-1
	=
	\frac{1}{2\sqrt{\Phi_p^{(0)}}}
	+
	\frac{1}{4\Phi_p^{(0)}}
	+
	\mathcal{O}\left([\Phi_p^{(0)}]^{-3/2}\right).
\end{equation}
Retaining only the leading term in this expansion, Eq.~(\ref{eq:transcendental_peak}) reduces to
\begin{equation}
	2\left(1-e^{-u_p}\right)-u_p
	\simeq
	\mathrm{OD}\left(2e^{-u_p}-1\right)^2 .
\end{equation}
This relation gives $\ln2-u_p=\mathcal{O}(\mathrm{OD}^{-1/2})$. To determine the next correction consistently, the $1/(4\Phi_p^{(0)})$ term in the Bessel-function expansion must also be retained. Expanding Eq.~(\ref{eq:transcendental_peak}) about $u_p=\ln2$ through order $\mathrm{OD}^{-1}$ then gives
\begin{equation}
	\label{eq:asymptotic_peak}
	t_{\mathrm{peak}}
	\simeq
	\frac{1}{\Gamma_{21}}
	\left[
	\ln2
	-
	\sqrt{\frac{1-\ln2}{\mathrm{OD}}}
	-
	\frac{1+\ln2}{2\,\mathrm{OD}}
	+
	\mathcal{O}\left(\mathrm{OD}^{-3/2}\right)
	\right].
\end{equation}

Equation~(\ref{eq:asymptotic_peak}) gives the formal large-OD limit within the spatially uniform, independent-decay analytical model,
\begin{equation}
	t_{\mathrm{peak}}
	\longrightarrow
	\frac{\ln2}{\Gamma_{21}} .
\end{equation}
Within this model, the peak time approaches the limiting value from below as OD increases. For $\Gamma_{21}\simeq2\pi\times6.067~\mathrm{MHz}$, the corresponding timescale is approximately $18.18~\mathrm{ns}$. It coincides with the time at which the independent-decay inversion $\bar{\sigma}_3^{(0)}(t)=2e^{-\Gamma_{21}t}-1$ crosses zero, corresponding to equal excited- and ground-state populations. At this crossing, the inversion-dependent collective gain term vanishes; beyond the crossing, it becomes absorptive as $\bar{\sigma}_3^{(0)}(t)$ becomes negative. The limiting value therefore reflects the single-atom radiative-decay timescale at which the positive-inversion gain window closes.

In the inset of Fig.~\ref{fig:fig2}(b), Eq.~(\ref{eq:asymptotic_peak}) is shown by the dashed line and compared with the spatially resolved result shown by the solid line. Over the intermediate-OD range, the analytical expression captures the overall OD dependence of the peak time, although a quantitative deviation from the spatially resolved result remains. At lower OD, the discrepancy partly reflects finite-$\Phi_p^{(0)}$ corrections omitted from the truncated asymptotic expansion. At higher OD, the breakdown of the spatially uniform inversion approximation becomes increasingly important, as the localized depletion near the output end advances the pulse peak (Sec.~\ref{subsec:cooperative_peak_time}), while the independent-decay treatment inherently neglects the macroscopic feedback of collective emission. The formal large-OD limit in Eq.~(\ref{eq:asymptotic_peak}) is therefore a limit of the reduced analytical model, whereas the spatially resolved system exhibits the reversal of the peak-time trend shown in Fig.~\ref{fig:fig2}(b).


\subsection{Predictive accuracy and limits of the uniform-inversion approximation}
\label{subsec:observable_dependent_limits}

To assess the predictive accuracy of the spatially uniform inversion approximation, we compare its calculated inversion dynamics and forward emission with spatially resolved stochastic benchmarks in Fig.~\ref{fig:fig3}. At $\mathrm{OD}=10$, the two emission profiles are in close agreement [Fig.~\ref{fig:fig3}(a)]. Furthermore, excellent quantitative agreement in both peak height and pulse shape persists at $\mathrm{OD}=50$ [Fig.~\ref{fig:fig3}(b)] and remains robust up to $\mathrm{OD}=200$ [Fig.~\ref{fig:fig3}(c)]. For $\mathrm{OD}=10$, $50$, and $200$, the inversion obtained within the uniform approximation remains close to the independent-decay trajectory $2e^{-\Gamma_{21}t}-1$ [Fig.~\ref{fig:fig3}(d)]. Consequently, Eq.~(\ref{eq:uniform_inversion_solution}) quantitatively predicts the absolute forward photon emission rate, not merely its normalized temporal profile, over the moderate-OD range examined here, although the extent of this agreement depends on the effective mode area and corresponding atom number.

This quantitative success stems from the distinct spatial roles played by the population inversion and the optical field. In an extended medium, forward superradiance arises from the propagation and collective amplification of microscopic polarization fluctuations. As established by the population balance in Eq.~(\ref{eq:inversion_decomposition}) and Fig.~\ref{fig:fig2}(c), the collective-depletion term $(2/N)I_{\mathrm{coll}}(t)$ remains subdominant to $2e^{-\Gamma_{21}t}$ through most of the buildup, consistent with the weak longitudinal inversion variations at $\mathrm{OD}=200$ seen in Fig.~\ref{fig:fig1}(b). Crucially, while our model homogenizes the population inversion, it retains longitudinal field propagation and the spatial dependence of atomic polarization through the modified Bessel kernel. The analytical solution therefore captures the collective buildup of the forward field without imposing spatial uniformity on the field or atomic polarization.

The peak time is also closely reproduced by the self-consistent uniform-inversion solution over the moderate-OD range. As shown in Sec.~\ref{subsec:analytical_peak_evaluation} and Eq.~(\ref{eq:asymptotic_peak}), its overall OD dependence is largely constrained by the reservoir-driven depletion of the inversion that limits the positive-inversion gain window. The additional independent-decay approximation captures this overall trend, although quantitative deviations remain, as shown in the inset of Fig.~\ref{fig:fig2}(b). This agreement, however, does not extend to arbitrarily high OD. As shown in Fig.~\ref{fig:fig2}(b), the spatially resolved peak-time trend eventually reverses, with the peak time decreasing as OD increases further. In this high-OD regime, the strongly amplified field rapidly depletes the local inversion near the exit face, consistent with the earlier peak of the collective emission pulse. Capturing this localized depletion and the resulting peak-time reversal requires a spatially resolved treatment.


\subsection{OD scaling of superradiance}
\label{sec:scaling_behavior}

\begin{figure}[t]
	\centering
	\includegraphics[width=\columnwidth]{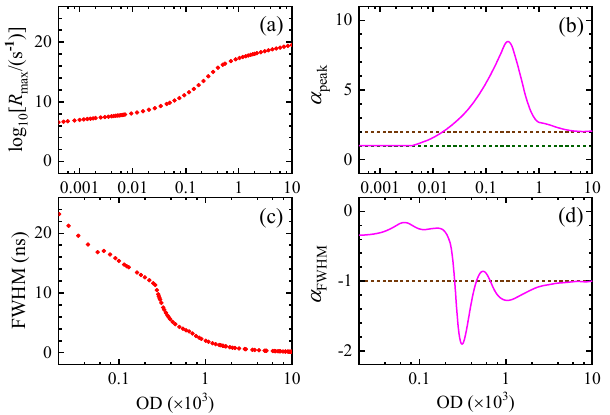}
	\caption{Scaling dynamics of forward superradiance obtained from spatially resolved simulations up to $\mathrm{OD}=10^4$. (a) $\log_{10}[R_{\max}/(\mathrm{s}^{-1})]$ as a function of OD, where symbols represent log-transformed simulation data. (b) Local scaling exponent of the peak rate, $\alpha_{\mathrm{peak}}=\mathrm{d}\ln R_{\max}/\mathrm{d}\ln\mathrm{OD}$, featuring dashed reference lines at $\alpha_{\mathrm{peak}}=2$ (quadratic peak-rate scaling) and $\alpha_{\mathrm{peak}}=1$ (linear low-OD scaling). (c) FWHM of the primary burst as a function of OD. (d) Pulse width scaling exponent $\alpha_{\mathrm{FWHM}}=\mathrm{d}\ln(\mathrm{FWHM})/\mathrm{d}\ln\mathrm{OD}$ of the primary burst alongside the reference line $\alpha_{\mathrm{FWHM}}=-1$. In (b) and (d), the logarithmic exponents are evaluated by differentiating spline interpolations of the discrete simulation data.}
	\label{fig:fig4}
\end{figure}

To understand the global transition from independent spontaneous emission to strongly cooperative emission, we perform a systematic OD scaling analysis across the entire $\mathrm{OD}$ range up to $\mathrm{OD}=10^4$ using the 1D stochastic Maxwell--Bloch framework, as summarized in Fig.~\ref{fig:fig4}. We first estimate the atomic density at the upper end of this OD range. For the extended ensemble length $L=4\text{ mm}$, $\mathrm{OD}=10^4$ corresponds to an atomic number density of $n=\mathrm{OD}/(\sigma_0 L)\simeq8.6\times10^{12}\text{ cm}^{-3}$. Defining the resonant wavevector $k_0=2\pi/\lambda$, the microscopic spatial density parameter evaluates to $n k_0^{-3}\simeq 0.016\ll1$. The average interatomic separation $r_{\mathrm{avg}}\sim n^{-1/3}\simeq 0.49\ \mu\text{m}$ corresponds to $k_0r_{\mathrm{avg}}\simeq3.9$. These estimates suggest that near-field dipole--dipole coupling is relatively weak at typical interatomic separations. The scaling analysis therefore focuses on collective forward emission within the 1D propagation model, without explicitly including microscopic dipole--dipole interactions or collisional processes~\cite{Friedberg1973}.

The logarithm of the peak forward photon emission rate $R_{\max}$ is plotted in Fig.~\ref{fig:fig4}(a), showing a nonlinear increase with OD. To characterize its local growth rate across the OD range, we define the logarithmic scaling exponent
\begin{equation}
	\alpha_{\mathrm{peak}}
	\equiv
	\frac{\mathrm{d}\ln R_{\max}}{\mathrm{d}\ln\mathrm{OD}},
\end{equation}
as illustrated in Fig.~\ref{fig:fig4}(b). At low OD, the emission maximum occurs at $t=0$, so $R_{\max}=R(0)=\mathrm{OD}\Gamma_{21}/4$ and $\alpha_{\mathrm{peak}}\simeq1$, corresponding to linear scaling with $N$ at fixed sample geometry. As OD increases, a delayed maximum develops at $t_{\mathrm{peak}}>0$, and the local scaling exponent changes rapidly near the onset of cooperative buildup. As the system enters the moderate-OD regime, collective amplification strengthens the forward emission relative to spontaneous decay into other radiative modes. In this crossover regime, $\alpha_{\mathrm{peak}}$ rises above two before decreasing at higher OD. This overshoot occurs as a delayed cooperative burst develops during the crossover and does not represent a distinct asymptotic power law. At $\mathrm{OD}\gtrsim4\times10^3$, $\alpha_{\mathrm{peak}}$ remains close to two up to $\mathrm{OD}=10^4$, indicating a trend toward Dicke-like quadratic peak-rate scaling, $R_{\max}\propto N^2$, at fixed sample geometry. This behavior does not imply that the extended ensemble reaches the small-sample Dicke limit.

The OD-dependent increase in peak forward emission rate is accompanied by a systematic narrowing of the primary superradiant burst. We evaluate the full width at half maximum (FWHM) of the primary burst for $R_{\max}>2R(0)$, which ensures that the initial emission rate lies below half maximum, as displayed in Fig.~\ref{fig:fig4}(c). The FWHM decreases with increasing OD, showing a pronounced reduction in the crossover region where a well-developed delayed burst emerges, followed by continued narrowing at higher OD. The corresponding scaling exponent is
\begin{equation}
	\alpha_{\mathrm{FWHM}}
	\equiv
	\frac{\mathrm{d}\ln(\mathrm{FWHM})}{\mathrm{d}\ln\mathrm{OD}}.
\end{equation}
As shown in Fig.~\ref{fig:fig4}(d), the local width-scaling exponent varies nonmonotonically with OD. In the moderate-OD regime, the exponent decreases well below $-1$, reaching approximately $-2$. This transient steepening occurs during the crossover from weak collective buildup to a well-developed delayed burst and does not establish a separate asymptotic power law. At higher OD, the exponent approaches $-1$, consistent with the inverse-OD cooperative timescale $\tau_R=(\mathrm{OD}\,\Gamma_{21})^{-1}$ increasingly governing the duration of the main emission pulse. The high-OD behavior is consistent with Dicke-like quadratic growth of the peak emission rate, while the pulse width decreases approximately inversely with OD. Recent theory has established geometric-optical-depth-dependent bounds on state-optimized many-body emission rates for fixed-density ensembles and examined directional-detection scaling~\cite{Rusconi2026}. In comparison, the present analysis characterizes the forward-channel peak rate and pulse width dynamically attained from an initially inverted state as OD varies at fixed sample geometry.

	
\section{Low-excitation limit and OD-dependent forward-emission decay}
\label{sec:low_excitation}

In contrast to the fully inverted regime studied above, we now consider the linear low-excitation limit, in which the ensemble remains predominantly in the ground state and the inversion can be approximated as $\bar{\sigma}_3(t)\simeq-1$. We assume a spatially uniform weak excitation with an initial excited-state fraction $p_e\ll1$, vanishing mean polarization and no initial interatomic dipole correlations. The coupled atom-field equations then describe an absorption-dominated linear propagation problem. The weak excited-state population sets the overall magnitude of the initial local dipole correlation and hence of $R(0)=p_e\,\mathrm{OD}\Gamma_{21}/4$, but this factor cancels in the normalized rate $R(t)/R(0)$, so the normalized temporal response is independent of the initial excited-state population. The resulting solution provides an analytically tractable description of the OD-dependent short-time decay of the forward emission~\cite{Rehler1971,Scully2006,Scully2009}.

\begin{figure}[t]
	\centering
	\includegraphics[width=\columnwidth]{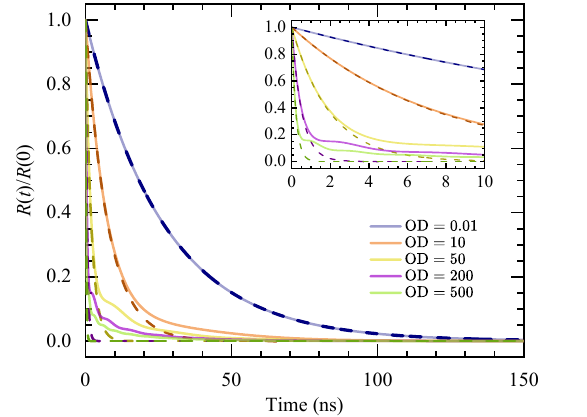}
	\caption{Normalized forward photon emission rate $R(t)/R(0)$ in the low-excitation regime for $\mathrm{OD}=500$, $200$, $50$, $10$, and $0.01$. Solid curves show the linear low-excitation solution [Eq.~(\ref{eq:low_excitation_full})], while dashed curves show the short-time exponential approximation in Eq.~(\ref{eq:low_excitation_decay}).}
	\label{fig:fig5}
\end{figure}

With $\bar{\sigma}_3(t)\simeq-1$, the propagation equations admit an analytical solution for each OD. Equation~(\ref{eq:Phi_definition}) then gives $\Phi(t)=-\mathrm{OD}\Gamma_{21}t/4$. Defining the dimensionless variable $\zeta=\sqrt{\mathrm{OD}\Gamma_{21}t}$ and using $I_0(i\zeta)=J_0(\zeta)$ and $I_1(i\zeta)=iJ_1(\zeta)$ in Eq.~(\ref{eq:uniform_inversion_solution}), the normalized solution becomes
\begin{equation}
	\label{eq:low_excitation_full}
	\frac{R(t)}{R(0)}
	=
	e^{-\Gamma_{21}t}
	\left[
	J_0^2(\zeta)+J_1^2(\zeta)
	\right].
\end{equation}
Since $\mathrm{d}[J_0^2(\zeta)+J_1^2(\zeta)]/\mathrm{d}\zeta=-2J_1^2(\zeta)/\zeta\leq0$, the emission rate decreases monotonically, giving $R_{\max}=R(0)=p_e\,\mathrm{OD}\Gamma_{21}/4\propto N$ at fixed sample geometry and $p_e$, even at high OD within the low-excitation limit. The Bessel-function factor retains the longitudinal propagation dynamics of the extended ensemble. Its initial decay becomes faster with increasing OD, reflecting the OD-dependent collective propagation of the forward emission in the weak-excitation regime. Because each curve is normalized by its initial value, Fig.~\ref{fig:fig5} compares the temporal decay profiles but does not show the OD dependence of the absolute forward photon emission rate.

In the short-time limit $\zeta\ll1$, the ordinary Bessel functions satisfy $J_0(\zeta)\simeq1-\zeta^2/4$ and $J_1(\zeta)\simeq\zeta/2$. Thus, $J_0^2(\zeta)+J_1^2(\zeta)=1-\zeta^2/4+\mathcal{O}(\zeta^4)$, and the normalized emission rate becomes
\begin{equation}
	\label{eq:low_excitation_decay}
	\frac{R(t)}{R(0)}
	\simeq
	\exp\left[
	-\Gamma_{21}
	\left(
	1+\frac{\mathrm{OD}}{4}
	\right)t
	\right].
\end{equation}
The corresponding short-time exponential decay rate of the forward-emission signal is therefore $\Gamma_{21}(1+\mathrm{OD}/4)$, which increases linearly with OD. The term $\Gamma_{21}$ gives the single-atom contribution to the initial decay, whereas the additional term $\mathrm{OD}\Gamma_{21}/4$ arises from the collective propagation dynamics and becomes increasingly important with OD. The numerical coefficient $1/4$ follows from the coupling and OD normalization of the present 1D model, while the OD-enhanced initial decay of the forward-emission signal is consistent with cooperative-emission behavior in extended atomic ensembles~\cite{Rehler1971,Scully2006,Scully2009}. A related OD-dependent temporal narrowing occurs in superradiant biphoton emission generated through spontaneous four-wave mixing in the weak-excitation regime, where an effective collective two-level response yields the same OD-dependent decay timescale~\cite{Chen2026}.

The full solution also captures the slower decay tail that emerges beyond the initial OD-enhanced regime. For $\zeta\gg1$, the leading asymptotic forms of the ordinary Bessel functions give $J_0^2(\zeta)+J_1^2(\zeta)\simeq2/(\pi\zeta)$, so that
\begin{equation}
	\frac{R(t)}{R(0)}
	\simeq
	\frac{2}{\pi\sqrt{\mathrm{OD}\Gamma_{21}t}}
	e^{-\Gamma_{21}t}.
\end{equation}
The rapid OD-enhanced initial decay therefore crosses over to a nonexponential Bessel-function tail with an algebraically decaying propagation factor multiplying $e^{-\Gamma_{21}t}$. This tail becomes increasingly prominent relative to the short-time exponential approximation as OD increases.

The exponential approximation applies only at early times satisfying $\mathrm{OD}\Gamma_{21}t\ll1$, so its validity interval decreases with increasing OD. Accordingly, this OD-dependent rate characterizes only the initial short-time decay of the forward-emission signal and is not a global exponential decay constant or a decay rate of the excited-state population. Its linear dependence on OD strictly describes the low-excitation temporal decay rate and remains distinct from the OD scaling of the peak forward-emission rate in the fully inverted nonlinear regime.


\section{Conclusion}

We have presented an open-system quantum theory of forward superradiance in an extended 1D two-level ensemble based on coupled Heisenberg--Langevin and Maxwell propagation equations. For a fully inverted initial state with zero mean polarization, the emission is initiated by nonzero local dipole correlations in the operator-based analytical formulation without introducing a prescribed tipping angle or stochastic seed, with field-mediated collective interactions producing a delayed superradiant burst at sufficiently high OD.

Our central result is an analytical expression for the absolute forward photon emission rate, obtained by treating only the population inversion as spatially uniform while retaining longitudinal field propagation and spatially dependent atomic polarization. Across the moderate-OD regime, this expression, combined with self-consistent mean-inversion dynamics, quantitatively predicts both the temporal profile and the absolute amplitude obtained from spatially resolved stochastic simulations. This agreement establishes the quantitative predictive capability of the reduced analytical model despite its uniform-inversion approximation.

The OD-dependent peak time further illustrates the validity and limitations of this approximation. Within the spatially uniform, independent-decay model, the analytical peak time approaches $\ln2/\Gamma_{21}$ from below with a leading correction proportional to $1/\sqrt{\mathrm{OD}}$, reflecting the single-atom timescale at which the independent-decay inversion crosses zero and the positive-inversion gain window closes. At higher OD, strong longitudinal inversion gradients accompany the reversal of the peak-time trend, revealing the limitations of the spatially uniform inversion approximation. The spatially resolved scaling analysis reveals a nonmonotonic local peak-rate exponent exceeding two at intermediate OD before approaching two at the highest OD studied, while the pulse-width scaling exponent approaches $-1$.

Finally, in the low-excitation regime, the theory reduces to a linear absorption-dominated propagation problem and yields a short-time exponential decay rate of the forward-emission signal, $\Gamma_{21}(1+\mathrm{OD}/4)$, that increases linearly with OD. In contrast to the fully inverted regime, the peak emission rate scales as $R_{\max}\propto\mathrm{OD}$ at fixed initial excited-state fraction and sample geometry, without a delayed cooperative burst. The full solution further exhibits a nonexponential Bessel-function tail at later times, distinguishing this initial decay rate from a global exponential decay constant. Together, these results connect quantum-fluctuation initiation, longitudinal propagation, and inversion depletion, providing a quantitative understanding of superradiant burst dynamics and the limits of reduced analytical descriptions.

	
\begin{acknowledgments}
	
This work was supported by the National Science and Technology Council of Taiwan under Grant Nos. 114-2112-M-006-007 and 115-2112-M-006-001. Additional support was provided through the Center for Quantum Science and Technology under the Higher Education Sprout Project of the Ministry of Education, Taiwan.
	
\end{acknowledgments}

\vspace*{0.8\baselineskip}

\section*{Data Availability}

The data supporting the findings of this study are available from the corresponding author upon reasonable request.


	
\end{document}